\documentclass[conference]{IEEEtran}
\IEEEoverridecommandlockouts
\usepackage{cite}
\usepackage{amsmath,amssymb,amsfonts}
\usepackage{graphicx}
\usepackage{textcomp}
\usepackage{xcolor}

\usepackage{multirow} % my addition in tables
\usepackage{listings} % my addition for sql code
\usepackage{accents} % my addition for undercurrent accent
\usepackage{txfonts} % my addition for the mathematical term product \varprod
\usepackage{url} % my addition for url
\usepackage{varwidth} % my addition for varwidth
\usepackage{colortbl} % my addition for cell color table
\usepackage{hhline} % my addition for double line in table

\usepackage[noend]{algpseudocode}  %%% to remove endif, endfor,... 
\usepackage[Algorithm]{algorithm}

\newtheorem{definition}{Definition}

\def\BibTeX{{\rm B\kern-.05em{\sc i\kern-.025em b}\kern-.08em
		T\kern-.1667em\lower.7ex\hbox{E}\kern-.125emX}}

\definecolor{codegreen}{rgb}{0,0.6,0}
\definecolor{codegray}{rgb}{0.5,0.5,0.5}
\definecolor{codepurple}{HTML}{C42043}
\definecolor{codeblue}{HTML}{0000FF}
\definecolor{backcolour}{HTML}{F2F2F2}
\definecolor{bookColor}{cmyk}{0,0,0,0.90}  
\color{bookColor}

\lstdefinestyle{mystyle}{
	backgroundcolor=\color{backcolour},   
	commentstyle=\color{codegreen},
	keywordstyle=\color{codeblue},
	numberstyle=\numberstyle,
	stringstyle=\color{codepurple},
	basicstyle=\footnotesize\ttfamily,
	breakatwhitespace=false,
	breaklines=true,
	captionpos=b,
	keepspaces=true,
	numbers=left,
	numbersep=10pt,
	showspaces=false,
	showstringspaces=false,
	showtabs=false,
}
\newcommand\numberstyle[1]{%
	\footnotesize
	\color{codegray}%
	\ttfamily
	\ifnum#1<10 0\fi#1 |%
}

\begin{document}
	%Faculty of Computers and Artificial Intelligence, 
	\title{Correlating Unlabeled Events at Runtime}
	\author{\IEEEauthorblockN{Iman M. A. Helal}
		\IEEEauthorblockA{\textit{Information Systems Department} \\
			\textit{Fac. of Computers and AI, Cairo University}\\
			Giza, Egypt\\
			0000-0001-8434-7551} 
		\and
		\IEEEauthorblockN{Ahmed Awad}
		\IEEEauthorblockA{\textit{Information Systems Department} \\
			\textit{Fac. of Computers and AI, Cairo University}\\
			Giza, Egypt \\
			0000-0003-1879-1026}
	}
	\maketitle
	
	\begin{abstract}
		Process mining is of great importance for both data-centric and process-centric systems. Process mining receives so-called process logs which are collections of partially-ordered events. An event has to possess at least three attributes, case ID, task ID and a timestamp for mining approaches to work. When a case ID is unknown, the event is called unlabeled. Traditionally, process mining is an offline task, where events are collected from different sources are usually manually correlated. That is, events belonging to the same instance are assigned the same case ID. With today's high-volume/high-speed nature of, e.g., IoT applications, process mining shifts to be an online task. For this, event correlation has to be automated and has to occur as the data is generated. 
		
		In this paper, we introduce an approach that correlates unlabeled events at runtime. Given a process model, a stream of unlabeled events and other information about task duration, our approach can induce a case identifier to a set of unlabeled events with a trust percentage. It can also check the conformance of the identified cases with the process model. A prototype of the proposed approach was implemented and evaluated against real-life and synthetic logs.
				
	\end{abstract}
	
	
	\section{Introduction}
	Organizations can present their work as a set of procedures and their execution constraints. These procedures can be modeled as business processes, where each process is composed of a set of activities to serve a business goal. A business process execution generates a stream of events. The collection of such events is called execution logs. Each event is expected to convey valuable information for further analysis, monitoring and process enhancement approaches. These approaches assume the correlation of the event to a specific case identifier. However, this level of event maturity~\cite{ProcessMiningBook2016}  is guaranteed only when processes are executed within Process-Aware Information Systems or via Business Process Management Systems.
	
	In real life, the execution of business processes is mostly unmanaged. That is, there is no central orchestration (execution) engine that can correlate the events and ensure that all information is recorded. This leads to \emph{missing} information in the generated events. Whenever the case identifier is missing, this is called an unlabeled event~\cite{Ferreira2009}. Some research efforts have been devoted to the problem of unlabeled event correlation~\cite{Ferreira2009,Bayomie2015,Pourmirza2015,BayomieAE16,AbbadAndaloussi2018}. However, these approaches lend themselves to the batch (offline) processing of event logs, rather than online processing of events.
	
	Process mining~\cite{ProcessMiningBook2016} is a technique that analyzes such generated process events in order to : 1) discover (re-engineer) process models, 2) check conformance of execution traces to predefined process models, or 3) enhance existing models based on insights gained from log analysis. In general, process mining techniques assume the existence of at least three pieces of information within each event: Case ID, activity ID and the timestamp. The first correlates events executed with the same process instance, the second correlates events for the same task and the last one imposes order on the events.
	
	In general, process mining is done in an \emph{offline} mode. That is, logs are collected, correlated, processed after the fact. As a preprocessing, correlation of unmanaged events is done manually or semi-automatically. A number of offline auto-correlation approaches have been developed~\cite{Bayomie2015,BayomieAE16,BayomieCRM19}. With the advancement in data generation where the velocity of data puts emphasis on processing data as it is generated, e.g., the case of IoT applications~\cite{VitaliP16,hemmer:hal-02402986}, process mining techniques need to be extended to support \emph{online} processing. In a streaming setting, an auto-correlation has to provide 1) in-memory processing, 2) an incremental element-by-element processing, 3) ability to produce correlations immediately, that is no need to wait for the end of the stream, as it usually won't end, 4) possibility to provide immediate but probabilistic (approximate) results.
	
	In this paper, we contribute a runtime correlation approach on unlabeled streams of events. We build upon and enhance our previous work~\cite{Helal2015} by correlating each incoming event to its case upon arrival in near real-time. Each correlated (i.e. labeled) event has a probability reflecting the level of confidence about being a member of the specified case. An event may fail to correlate to any of the existing cases, which implies a deviation in the executed log from the original model. The contributions in this paper compared to our previous work are 1) performance and accuracy enhancement, 2) supporting both cyclic and acyclic models, and 3) storing the data in an in-memory database and using its indexing techniques to accelerate the processing at runtime.
	
	The remainder of this paper is organized as follows: a summary of related work in event correlation in Section~\ref{sec:related:work}. Section~\ref{sec:proposed:approach} presents an overview of the approach along with foundational concepts and techniques. Implementation details, evaluation, and comparison with related approaches are discussed in Section~\ref{sec:results:discussion}. Finally, we conclude the paper in Section~\ref{sec:conclusion:future} with a critical discussion and an outlook on future work.

	\section{Related work} \label{sec:related:work}
	
	For the unmanaged business processes, researchers are facing the issue of an incomplete correlated event log, where incomplete data are most likely to happen. This can be a problem for both process mining and monitoring. In~\cite{Mukhi2010}, the authors modeled the uncertainty associated with events using a probabilistic provenance model and deduce the timestamp of an activity with its level of confidence and accuracy. Another type of missing information was addressed in\unskip~\cite{Rogge-Solti2013}. The authors are using constrained activity durations, similar to execution heuristics. Their work is presenting a predictive monitoring using a stochastic process model, to repair missing events in the log. They predict the remaining execution times based on last observed time manually. 
	
	In~\cite{Folino2012,VanderAalst2012}, a correlated log is required for conformance or performance checking. Hence, an uncorrelated log can be very difficult to handle without an intermediate step as in~\cite{Bayomie2015,Helal2015,BayomieCRM19}. However, the complexity and performance of the application in~\cite{Bayomie2015,Helal2015,BayomieCRM19} were unacceptable for either runtime environment where incoming events can come with high speed and huge amount, or while auditing in process mining.

	In~\cite{Dustdar2006}, the authors proposed a semi-automated system with minimal manual tuning for correlating event names with the process model activity name using cases. They are matching actual event names in the execution log to the more abstracted activity names in the process model for conformance checking. Web services are becoming one of the main sources for a rich event log~\cite{Dustdar2006}. The execution of web services may have missed important information, e.g. workflow and case identifiers, to help analyzing workflow execution log. This will need extra information about execution time heuristics to help with correlation. 
	
	Other researchers are addressing specifically the uncorrelated log problem and how to deal with it~\cite{Ferreira2009,Perez-Castillo2014}. In~\cite{Ferreira2009}, the authors introduced an Expectation-Maximization approach to estimate a Markov model from an uncorrelated event log. It finds a single solution most often with a local maximum of the likelihood function. It has some limitations, where handling loops and the existence of parallelism may lead to incorrectly correlating some events in the uncorrelated log. While in~\cite{Perez-Castillo2014}, authors generate a correlated log based on an intermediate uncorrelated log collected at execution time. 
	
	In~\cite{Herzberg2013}, the authors have presented a framework to correlate events generated from the manual process execution environment. They normalize events using extra data from the event log, in order to generate more information to correlate raw events. There are few researches addressing cyclic process models,~\cite{Leemans2015,BayomieAE16,BayomieCRM19}. These researches are breaking down the originally cyclic model to sub-graphs to easily analyze their loop entry points and detect the main structure of the process model. However, they cannot be used directly in runtime environment.
	
	In~\cite{VandenBroucke2014}, the authors have used a real-time monitoring for conformance checking. Their main goal was to aid the systems with time-critical nature to discover possible violations or deviations from the original model behavior. While in~\cite{VanZelst2017}, authors are using an incremental way in computing prefix-alignments to help in accelerating the process of online performance checking and enhancing memory efficiency. However, both approaches assume that the input stream of events is of good quality and is correlated to their respected cases.
	
	The authors in~\cite{Cheng2017} are targeting the large logs, and how to correlate them without consuming a lot of time using Spark. They propose a ``Rule Filtering and Graph Partitioning'' (RF-GraP) approach which correlate events in a distributed environment. They used filtering-and-verification principle as well as a graph partitioning approach to help filter and correlating events respectively. However, their main target is to correlate events to its missing attributes based on a set of correlation conditions. Moreover, they are working in an offline environment.
	
	An approach targeting large stream of events is stream-based abstract representation (S-BAR) architecture~\cite{VanZelst2018}. One of the reasons for incomplete event details or event abstraction is due to the anonymity, where businesses are prohibited by laws and regulations from storing these details. This approach uses the same concept of abstraction used in several process discovery techniques, e.g. Heuristic Miner and Inductive Miner. Their evaluation is based on memory usage and processing times for process discovery from the stream of events.
	
	In~\cite{DeMurillas2019}, the authors present a framework that focuses on the quality of event logs extracted from complex database schemas and large amounts of data. They recommend process views to the user and rank them by interests. Events are correlated based on case notion concept to get the traces and assess their interestingness. Also the authors in~\cite{BayomieCRM19} use the traces to assess the correctness of the offline correlation of the event log.
	
	\section{Proposed approach}\label{sec:proposed:approach}

\begin{figure}[tb]
		\centerline{\includegraphics[width=0.5\textwidth]{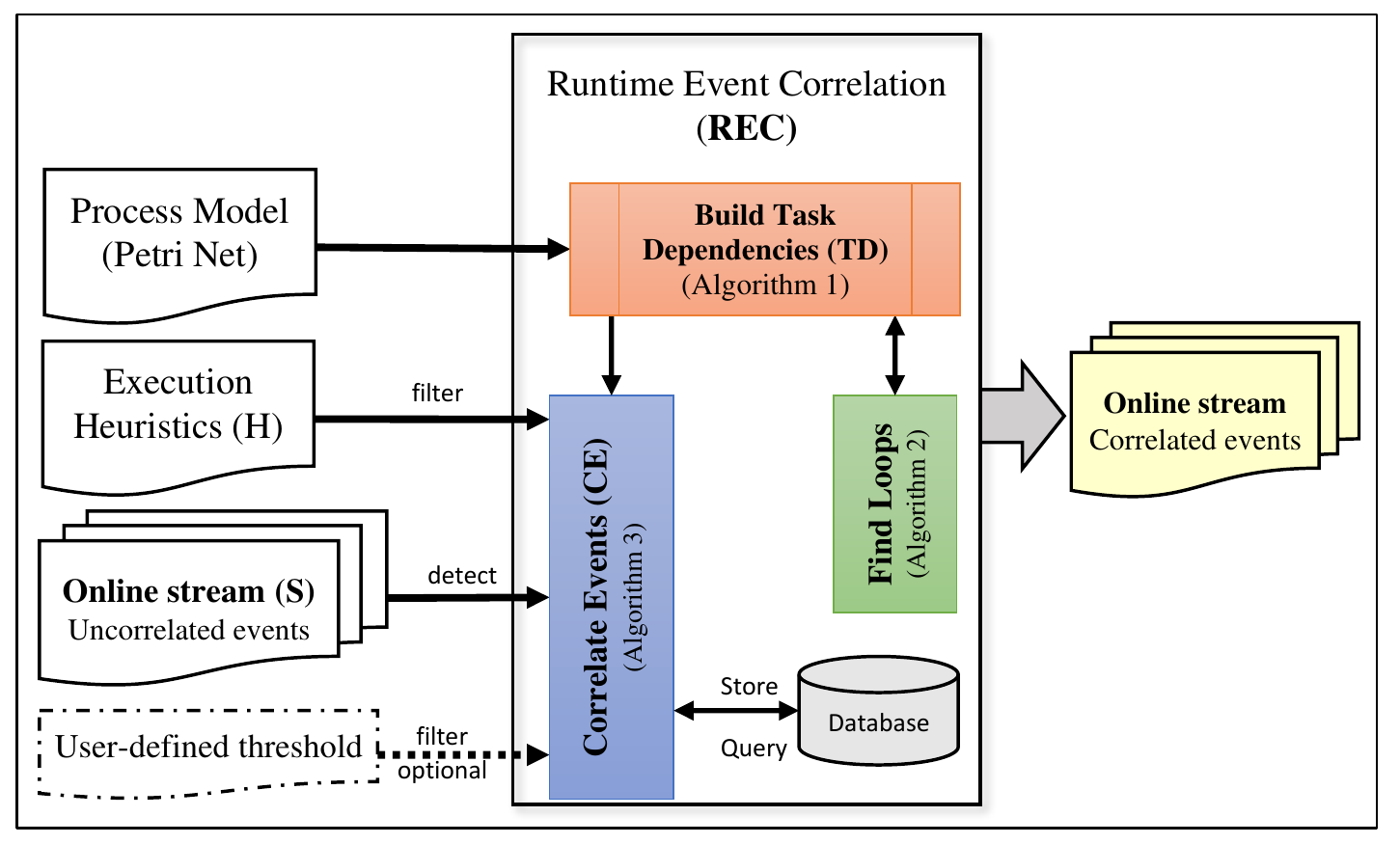}}
		\caption{{Approach overview}}
		\label{fig:approach:overview}
	\end{figure}

	The overview of Runtime Event Correlation (REC) approach is depicted in Fig.~\ref{fig:approach:overview}. It has three main inputs: 1) the process model, 2) process execution heuristics, and 3) the stream of uncorrelated events. As an output, for each input uncorrelated event, REC produces a set of correlated events due to the inherent uncertainty, as a single uncorrelated event might belong to more than one case with different trust percentages. This output can be filtered with an \textit{optional} user-specified threshold.

	The process model is the first input to our approach. We assume that it is presented as a workflow net~\cite{VanderAalst2004}, $PN(P,T,F) $, where P is the set of places, T is the set of transitions (a-k-a activities A), and the F is the set of Flows i.e. $F\subset(P\times T)\;\;\cup(T\times P) $. Moreover, we define the auxiliary sets on transition and place levels as $\bullet  t=\{p\in P:(p,t)\in F\} $, $t\bullet =\{p\in P:(t,p)\in F\} $, $\bullet  p=\{t\in T:(t,p)\in F\} $, and $p\bullet =\{t\in T:(p,T)\in F\} $.
	
	The second input is the execution heuristics. It can be extracted from a sample execution of events, to capture and predict activity durations. This is possible in a managed environment, where business processes are well-defined and are fully automated. Hence, there is no human interventions to add uncertainty to activity durations. However, in an unmanaged environment, an activity duration will need more investigation. In small systems, an expert can help identify activity durations. However, this is not feasible or applicable for large and diverse systems. Although there are different systems that are Process-Aware Information System (PAIS), but still a few activities can be done through human interactions; e.g. logistics, and health-care systems. In \cite{Gon2016}, the authors identified a stochastic technique to capture user activities durations in an unmanaged environment, i.e. for low-level event logs.
	
	Our approach is composed of three main components, Build Task Dependencies, Find Loops, and Correlate Events (CE). The first component analyzes the process model structurally and derives the different causal dependencies for each task. This step is done once per process model change and generates task dependencies (TD). TD is a structure where each task has a set of predecessor tasks. In order to handle different types of cyclic behaviors, such as structured or unstructured loops of any length, it calls Find Loops component.
	
	Finally, CE uses TD to check the inter-dependencies between the different tasks. It also uses the execution heuristics H to infer a case identifier for the incoming stream of uncorrelated events. With the aim of managing the huge number of events, CE uses a database engine. It stores the correlated event instances for each incoming uncorrelated event to accelerate the process of correlation using indexing. It indexes each activity in the set of Activities from the business process model and each newly detected case, as \textit{Activities} and \textit{Cases} indexes, respectively. 
	
	REC can handle noise in the incoming stream of events. The source of noise can be either incorrect event detection which may cause violations for compliance monitoring, a deviation from the original process model (i.e. non-conforming trace), or a missing scenario in the process model due to incomplete execution or a rare occurrence \cite{VanderAalst2004}. In case of any detected noisy events, a flag is raised at runtime to be further addressed. 
	
	\subsection{Running Example}
	\label{sub:sec:example}
	
\begin{figure}[tb]
		\centerline{\includegraphics[width=0.5\textwidth]{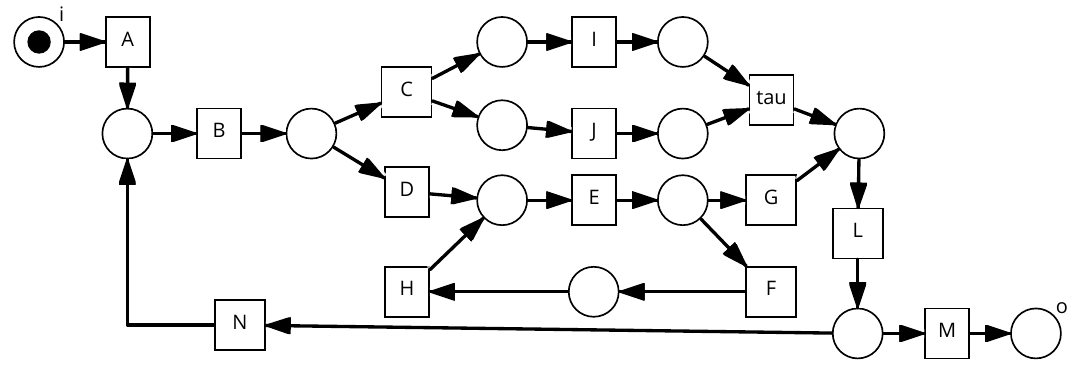}}
		\caption{{A patient physical examination business process (PN)\unskip~\protect\cite{PhysicalExamExample}}}
		\label{fig:physical:example}
	\end{figure}

	In a clinic, a patient can do physical examination \cite{PhysicalExamExample}, fig.~\ref{fig:physical:example}. The process starts with adding patient details (A), then check for medical history (B). The physician can do wellness check-up (C) by checking patient's current medications (I), and checking for old surgeries (J) in order to complete the wellness check-up (tau). Otherwise, the patient specifies the symptoms (D). Per each symptom, the physician checks the symptom's details (E) and do suitable check (e.g. draw blood, scan) (F) and analyzes the results of this check (H). Then, the physician finalizes symptoms check-up (G). The patient medical history is updates (L) and the physician checks if there are any emerging complications (N) to repeat the procedure. Finally, the patient finalizes the medical check-up (M). Note that \textit{tau} is a silent transition with no execution time.
	
	Table.~\ref{tbl:heurisitic:data} shows the execution heuristics for the running example, where all units are in minutes. Tables~\ref{tbl:uncorrelated:1},\ref{tbl:uncorrelated:2} show snapshots of the stream of uncorrelated events (S). S can have details regarding the life cycle of the activity and the resource responsible, or these details can be unavailable (low-level log).

	\begin{table}[bt]
		\caption{Heuristic Data (H)}
		\label{tbl:heurisitic:data}
		\ignorespaces 
		\begin{center}
			\begin{tabular}{|@{ }c@{ }|@{ }c@{ }|@{ }c@{ }|@{ }c@{ }|@{ }c@{ }|@{ }c@{ }|@{ }c@{ }|@{ }c@{ }|@{ }c@{ }|@{ }c@{ }|@{ }c@{ }|@{ }c@{ }|@{ }c@{ }|@{ }c@{ }|}
				\hline
				\textbf{Activity} & \textbf{A} & \textbf{B} & \textbf{C} & \textbf{D} & \textbf{E} & \textbf{F} & \textbf{G} & \textbf{H} & \textbf{I} & \textbf{J} & \textbf{L} & \textbf{M} & \textbf{N} \\ \hline
				\textbf{Min} & 1 & 1 & 1 & 1 & 1 & 2 & 1 & 3 & 1 & 3 & 2 & 1 & 1 \\ \hline
				\textbf{Avg} & 1 & 3 & 2 & 1 & 4 & 2 & 2 & 3 & 4 & 4 & 7 & 5 & 1 \\ \hline
				\textbf{Max} & 1 & 4 & 2 & 1 & 7 & 2 & 3 & 3 & 7 & 4 & 11 & 9 & 1 \\ \hline
			\end{tabular}%
		\end{center}
	\end{table}

	\begin{table}[bt]
		\caption{{Stream of Uncorrelated Events (S) with implicit Lifecycle=Completed}} 
		\label{tbl:uncorrelated:1}
		\def\arraystretch{1}
		\ignorespaces 
		\scriptsize
		\begin{center}
			
			\begin{tabular}{@{ }c@{ }@{ }c@{ }@{ }c@{ }||@{ }c@{ }@{ }c@{ }@{ }c@{ }}
				\hline
				CaseID & Activity & Timestamp & CaseID & Activity & Timestamp \\
				\hline
				- & A & 2019-6-16 11:55:01 & - & E & 2019-6-16 11:55:17 \\
				- & A & 2019-6-16 11:55:02 & - & H & 2019-6-16 11:55:18 \\
				- & B & 2019-6-16 11:55:03 & - & E & 2019-6-16 11:55:19 \\
				- & C & 2019-6-16 11:55:04 & - & G & 2019-6-16 11:55:20 \\
				- & A & 2019-6-16 11:55:05 & - & L & 2019-6-16 11:55:21 \\
				- & B & 2019-6-16 11:55:06 & - & L & 2019-6-16 11:55:22 \\
				- & D & 2019-6-16 11:55:07 & - & N & 2019-6-16 11:55:23 \\
				- & J & 2019-6-16 11:55:08 & - & B & 2019-6-16 11:55:24 \\
				- & B & 2019-6-16 11:55:09 & - & M & 2019-6-16 11:55:25 \\
				- & D & 2019-6-16 11:55:10 & - & C & 2019-6-16 11:55:26 \\
				- & I & 2019-6-16 11:55:11 & - & I & 2019-6-16 11:55:27 \\
				- & E & 2019-6-16 11:55:13 & - & M & 2019-6-16 11:55:28 \\
				- & G & 2019-6-16 11:55:14 & - & J & 2019-6-16 11:55:29 \\
				- & F & 2019-6-16 11:55:15 & - & L & 2019-6-16 11:55:31 \\
				- & L & 2019-6-16 11:55:16 & - & M & 2019-6-16 11:55:32 \\
				\hline
			\end{tabular}%
		\end{center}
	\end{table}

	\begin{table}[bt]
		\caption{{Stream of Uncorrelated Events (S) with Lifecycle = \tt}\{Started,Completed\}} 
		\label{tbl:uncorrelated:2}
		\def\arraystretch{1}
		\ignorespaces 
		\scriptsize
		\begin{center}
			\begin{tabular}{ccclc}
				\hline 
				CaseID & Timestamp & Activity & LifeCycle & Resource \\ \hline
				
				― & 2019-06-16 11:55:00 & A & Started & Noah \\  
				― & 2019-06-16 11:55:01 & A & Completed & Noah \\  
				― & 2019-06-16 11:55:01 & A & Started & Sam \\  
				― & 2019-06-16 11:55:02 & A & Completed & Sam \\  
				\dots & \dots & \dots & \dots & \dots \\  
				
				― & 2019-06-16 11:55:04 & A & Started & Adam \\  
				― & 2019-06-16 11:55:05 & A & Completed & Adam \\  
				― & 2019-06-16 11:55:05 & B & Started & System \\  
				― & 2019-06-16 11:55:06 & B & Completed & System \\  
				\dots & \dots & \dots & \dots & \dots \\  
				\hline
				
			\end{tabular}
		\end{center}
	\end{table}
	
	%%%%%%%%%%%%%%%%%%%%%%%%%%%%%%%%%%%%%%%%%%%%%%%%%%%%%%%%%%%%%%%%%%%%%%%%
	\subsection{Building Task Dependencies component}
	\label{sub:sec:build:TD}
	
	The output of this component acts as a preprocessing step for correlating a stream of events to their respective cases. Algorithm~\ref{algo:build:TD} uses an anomaly-free workflow net, i.e., both deadlock and livelock free~\cite{VanDerAalst2011}. If the input net is anomalous, then our algorithm might generate faulty results.

	Task Dependency uses the concept of a \emph{dependency graph}~\cite{VanderAalst2004,Weijters2011,Greco2012}. It explains the causality between activities. Each node represents an activity, and each directed edge represents causal dependency between activities. Edges are annotated with the frequency of occurrence and/or confidence or certainty of causality between those nodes (as in Disco tool \cite{Rozinat2012}). A \emph{directed graph} or \emph{digraph} $G=(V,E)$ is where $V$ is a finite set of \emph{vertices} or nodes, and $E \subseteq (V \times V )$ is a set of directed \emph{edges}~\cite{Diestel2010}.  A \emph{dependency graph} $DG$ of the graph $G$ uses labeled edges $E \subseteq (V \times V \times L)$ where $L$ is a set of labels to describe the dependencies between two nodes. 
	
	These annotations do not explain any semantics of process model execution. For example, $E=\{(v_1,v_2, l_1),$ $(v_2,v_3,l_2),$ $(v_1,v_3,l_3),\dots\}$ shows that there is a dependency from $v_3$ on one hand and  $v_1$ and $v_2$ on the other hand. However, it is unclear whether $v_3$ requires both, only one, or at least one of them. This representation is used in most process mining techniques to represent fuzzy models; e.g. ProM \cite{VerbeekBDA10} and Disco. In~\cite{Weijters2011}, a causal net (C-net) was presented to illustrate the input and output sets for each activity. 
	
	We use the dependency graph and C-net to clarify the causalities between tasks with extra semantics. Task dependency (cf. Definition~\ref{def:TaskDependencies}) links each activity with its possible predecessors/dependencies. It supports well-structured and unstructured (i.e. arbitrary) loops~\cite{Russell2016}.
	
	\begin{definition} 	[Task Dependency]  \label{def:TaskDependencies}%\\
		Let $A$ be a finite set of \emph{activities} in a business process model. Task dependency $TD : A \longrightarrow \mathcal{P}(\mathcal{P}(A))$, where $\mathcal{P}$ is the power set.
	\end{definition}
	
	For an activity $a\in A$, $TD(a) = \{S_1, S_2, \dots, S_n\}$. It is necessary to observe events corresponding to \emph{every} element of any $S_i$ before observing an unlabeled event of $a$ in order to correlate that event with the case in which events of $S_i$ were observed. In addition to task dependency, we identify activities that act as loop entry points in the workflow net. So, we define $isLoopEntry : A \longrightarrow \{True,False\}$.

	Algorithm~\ref{algo:build:TD} describes how the task dependency is derived. For each activity $t \in T$, the size of $\bullet t$ is  checked. In case there is only one input place $p$ for $t$, Algorithm~\ref{algo:build:TD} iterates over the input places of $p$ and appends each input transition $x$ in a separate set to $TD(t)$, see line \ref{line:build:TD:loop:single}. If on the other hand, $t$ has more than one input place, i.e. it is a synchronization node, for each $p \in \bullet t$ we get the set of preceding transitions and store them in set $M$, see line \ref{line:build:TD:store:M}. After each iteration $M$ is added as an element in set $D$, see line \ref{line:build:TD:store:D}. Transitions within $M$ represent exclusive transition. However, transitions from $M_1, M_2\in D$ are concurrent. That is, for $t$, it has to await any of the possible combinations across elements of sets $M_1 \dots M_{|D|}\in D$. This is represented by the non-Cartesian products of such sets, see line~\ref{line:non:cartesian:product}. 
	In line~\ref{line:identify:loop:entry:nodes}, Algorithm~\ref{algo:graph:traversal} is invoked to identify the loop entry nodes in the model. Starting from line~\ref{line:clean:tau}, it checks the dependencies of tasks and replaces dependencies on silent transitions with their predecessors so that all dependencies are on observable activities.

	%%%%%%%%%%%%%%%%%%%%%%%%%%%%%%%%%%%%%%%%%%%%%%%%%%%%%%%%%%%%%%%%%%%%%%%%%
	\begin{algorithm}[bt]\caption{Build Task Dependencies}\label{algo:build:TD}
		\scriptsize
		\begin{algorithmic}[1]
			\Require $G=PN(P,T,F)$ ~~~~//Process model as workflow net 
			\Ensure $TD$ //Task Dependencies for all activities $A$ without silent transitions (Def.~\ref{def:TaskDependencies})
			\State $tempTD$ // empty hashmap used for calculations
			\State $TD$ = new $TD$()	// task dependencies without silent transitions
			\ForAll {$(t \in T)$} 
			\If {$|\bullet t|==1$}
			\ForAll {$( x \in \bullet p \mid p\in \bullet t)$} \label{line:build:TD:loop:single}
			\State $tempTD[t].append(\{x\}) $
			\EndFor
			\ElsIf {$|\bullet t| >1$} ~~ // parallel transitions case
			\State $S = \{\}$
			\State $M = \{\}$
			\ForAll  {$(p \in \bullet t)$}
			\ForAll {$(x \in \bullet p)$}
			\State $M = M \cup \{x\} $ \label{line:build:TD:store:M}
			\EndFor
			\State $D = D \cup \{M\}$ \label{line:build:TD:store:D}
			\EndFor
			\State \label{line:non:cartesian:product} $tempTD[t] = \underaccent{\bullet}{\varprod}_{i=1}^{|D|} M_i$\footnote{$A \underaccent{\bullet}{\times} B=\{\{a,b\}: a \in A~and~b \in B\}$ is the non-Cartesian product of two sets A and B.} %\footnotemark 
			\EndIf
			\EndFor
			\State $TD = find\_loops(G, tempTD)$~~~~~~~~~//(cf. Algorithm~\ref{algo:graph:traversal})  \label{line:identify:loop:entry:nodes}	
			\ForAll {$(t \in T)$} ~~~~~~~~~~~~~~~~~~~~~~~~~~~~~~// `t' represents transition name \label{line:clean:tau}
			\If {$(not~isSilent(t))$}~~~~~~~// Is current transition not a silent transition?
			\ForAll {$(S \in TD(t))$} 
			\If {$(\exists x \in S \wedge isSilent(x))$} ~~~~~~~~~~// $x$ represents silent transition \label{alg:build:TD:clean:start}
			\State $S = S \setminus \{x\}$
			\State $S = S \cup TD(x)$  ~~~~~~~~~~// replace $x$ with its original dependencies \label{alg:build:TD:clean} 
			\EndIf \label{alg:build:TD:clean:end}
			\EndFor
			\EndIf 
			\EndFor
		\end{algorithmic}
	\end{algorithm}
	
	\footnotetext{$A \underaccent{\bullet}{\times} B=\{\{a,b\}: a \in A~and~b \in B\}$ is the non-Cartesian product of two sets A and B.}
	%%%%%%%%%%%%%%%%%%%%%%%%%%%%%%%%%%%%%%%%%%%%%%%%%%%%%%%%%%%%%%%%%%%%%%%%%
	%%%%%%%%%%%%%%%%%%%%%%%%%%%%%%%%%%%%%%%%%%%%%%%%%%%%%%%%%%%%%%%%%%%%%%
	\subsection{Find Loops component} \label{sub:sec:FL}

	Our approach detects which activity nodes are the entry points of the loop behavior without restructuring the original process model. Also, it does not need to break-down the main process model into sub-structures to separate and detect the cyclic paths \cite{Pourmirza2015,BayomieAE16}. 
	
	Algorithm~\ref{algo:graph:traversal} has been abstracted to clarify how loop entries are flagged. It has a dummy list $traversed$ to check and count the number of times a node was checked, see line \ref{line:alg:graph:traversal:traversed}. Line~\ref{line:alg:graph:traversal:recursive} starts the graph $G$ traversal for neighbor nodes using the depth-first search mechanism. Then, line~\ref{line:alg:DFS:traversed:once} ensures that no node is visited more than twice. If the current checked node ($ch\_node$) was revisited, see line~\ref{line:alg:DFS:traversed:revisit}, then that $node$ is a loop entry. It is flagged on line~\ref{line:alg:graph:traversal:traversed:twice}, then added to $loopEntries$ list. 
	
	%%%%%%%%% Self written, as footnote doesnot work with algorithmic correctly
	%\footnotetext{$A \underaccent{\bullet}{\times} B=\{\{a,b\}: a \in A~and~b \in B\}$ is the non-Cartesian product of two sets A and B.}

	%%%%%%%%%%%%%%%%%%%%%%%%%%%%%%%%%%%%%%%%%%%%%%%%%%%%%%%%%%%%%%%%%%%%%%%%%
	\begin{algorithm}[bt]\caption{Find Loops}\label{algo:graph:traversal}
		\scriptsize
		\begin{algorithmic}[1]
			\Require $G=(V,E)$ ~~~~~~~~~~~//Connected graph representing original model PN (P,T,F) 
			%\Require $tempTD$ ~~~~~ // a hashmap with initial predecessors extracted from PN
			\Ensure $loopEntries$ ~~~~~//List of loop entries in the model
			\State $traversed = []$ //hashmap (key,value); key:visited node, value: number of visits
			\State $counter = 0$
			\ForAll {$(node \in V)$}
			\State $ch\_node = node$
			\State $counter += 1$
			\State $traversed.clear()$ ~~~// ensure a clean list for Depth-First Search of $ch\_node$ \label{line:alg:graph:traversal:traversed}
			\If {$(node \notin traversed.keys())$}
			\State $traversed[node] = 1$ \label{line:traversed:node:increase}
			\State $neighbour\_nodes =\{v \mid v \in V \wedge (node, v) \in E\}$
			\ForAll {($n \in neighbour\_nodes $)} ~~~~~~//take a neighboring node  \label{line:alg:graph:traversal:recursive}
			\If {($n \notin traversed.keys()$)}: ~//Is neighbor node already visited?\label{line:alg:DFS:traversed:once}% \begin{varwidth}[t]{\linewidth}  \end{varwidth}\vspace{1mm}
			\State $traversed[n] = 1$ 
			\State $neighbour\_nodes =\{v \mid v \in V \wedge (n, v) \in E\}$
			\State $continue;$ \label{line:alg:DFS:traversed:recursive}
			\ElsIf {($n == ch\_node$)}~~~~//revisit same observing node\label{line:alg:DFS:traversed:revisit} 
			\State $traversed[node] =traversed[node] +1$
			\If {$(traversed[node]==2)$}~~~//possible loop entry?
			\State  $isLoopEntry(node) = True$~~~~~//flagging loop entry \label{line:alg:graph:traversal:traversed:twice}
			\State $loopEntries.add(node)$
			\EndIf
			\EndIf
			\EndFor
			
			\EndIf
			\EndFor
			
		\end{algorithmic}
	\end{algorithm}
	%\end{algorithmbis}
	
	%%%%%%%%%%%%%%%%%%%%%%%%%%%%%%%%%%%%%%%%%%%%%%%%%%%%%%%%%%%%%%%%%%%%%%%%%%

	Table.~\ref{tbl:TD} illustrates the task dependencies for the patient physical examination example, cf. Fig.~\ref{fig:physical:example}. For example, $TD(B)= \{\{A\},\{N\}\}$ represents possible dependencies for activity $B$. We find that $isLoopEntry(N)=True$, which represents that $N$ is part of a loop for activity $B$. However, activity $A$ is the main entry point for activity $B$. Hence, an event $B$ can only occur after an event $N$ in a case only if a previous occurrence of $B$ took place with an event $A$ as its predecessor. Also $TD(L)=\{\{G\},\{I,J\}\}$, where either $G$ or both activities $I$ and $J$ must have taken place in a case before the occurrence of activity $L$ in the same case. $TD(M)= \{\{L\}\}$ represents a causal dependency between activity $L$ and activity $M$, where activity $M$ depends on activity $L$'s occurrence. Finally, $TD(A) = \{\}$, indicates that activity $A$ is the start activity of the process model, as there is no dependencies on any other activities. At some scenarios, the business logic in a process model may require loop behavior for the start event. However, this scenario variation is not supported in our approach.

	\begin{table}[tb]
		\caption{Output Task Dependencies (TD)}
		\label{tbl:TD}
		\def\arraystretch{1}
		\ignorespaces 
		\begin{center}
			\begin{tabular}{|@{}c@{}|@{ }c@{ }|@{ }c@{ }|@{ }c@{ }|@{ }c@{ }|@{ }c@{ }|@{ }c@{ }|@{ }c@{ }|}%{|c|c|c|c|c|c|c|c|}
				\hline
				\textbf{Activity} & \textbf{A} & \textbf{B} & \textbf{C} & \textbf{D} & \textbf{E} & \textbf{F} & \textbf{G} \\ \hline
				\textbf{TD} & \{\} & \{\{A\},\{N\}\} & \{\{B\}\} & \{\{B\}\} & \{\{D\},\{H\}\} & \{\{E\}\} & \{\{E\}\} \\ \hhline{|========|}%\hline
				\textbf{Activity} & \textbf{H} & \textbf{I} & \textbf{J} & \textbf{L} & \textbf{M} & \textbf{N} & \textbf{\textit{Loop Entries}} \\ \hline 
				\textbf{TD} & \{\{F\}\} & \{\{C\}\} & \{\{C\}\} & \{\{G\},\{I,J\}\} & \{\{L\}\} & \{\{L\}\} & \textbf{\textit{D,G,H,I,J,N}} \\ \hline 
			\end{tabular}%
		\end{center}
	\end{table}

	%%%%%%%%%%%%%%%%%%%%%%%%%%%%%%%%%%%%%%%%%%%%%%%%%%%%%%%%%%%%%%%%%%%%%%
	\subsection{Correlate Events component} \label{sub:sec:EC}

	This component correlates and stores the event instances. Each incoming event has a corresponding activity from the process model. An event can be correlated to a case or more through different instances. We store event instances w.r.t their cases in the set ``CorrelatedEvents'' as in Definition \ref{def:CorrelatedEvents}. 
	
	\begin{definition}[Correlated Event] \label{def:CorrelatedEvents}
		Each event instance ($e_i$) has: ($timestamp$, $activity\_name$, $caseID$, $trust$, $lifecycle$, $resource$); where $caseID$, the correlated case of $e_i$. If $e_i$ is not correlated, then $caseID$ = $\bot$. Finally $trust$ is an assigned percentage for correlating $e_i$ to $caseID$, by default it is $\bot$. An event instance may have $lifecycle$ and/or $resource$ attributes, depending on the input stream format.
		
	\end{definition}
	
	At runtime, each incoming event $e$ has the main information: timestamp, activity\_name. In Definition \ref{def:CorrelatedEvents}, an event is cloned to different instances $e_i$, each of which is correlated to a different case based on TD(e.activity\_name); i.e. task dependencies $TD$ of the activity for that event. Each \emph{activity\_name} $\in$ \textbf{Activities} is as defined from the process model. The Activities set facilitates searching for all event instances and storing them into their respective ``activity''. Also, a set of \textbf{Cases}, based on \emph{caseID} attribute, helps in correlating the assigned events instances from the \emph{activity\_name} to their respective ``case''. The set of Cases is updated with each incoming start event; i.e. an event with no task dependencies (e.g. event ($1;A$)). 
	
	Correlate Events ($CE$) is responsible for finding possible allocations of each event instances for event $e$. First, it uses the output of Algorithm~\ref{algo:build:TD} to find the set of possible cases that contains $TD(e.activity\_name)$. For example, we refers to an event $e$ with timestamp $ts$ and life cycle $lifecycle$ as a simplified ($ts$;$activity\_name$;$lifecycle$). This event is part of a poor quality stream of events with either single value for the task life cycle (`completed') or (\{started, completed\}) (cf. Table.~\ref{tbl:uncorrelated:1},~\ref{tbl:uncorrelated:2} respectively). Some systems may provide details regarding the resource performing each task (cf. Table.~\ref{tbl:uncorrelated:2}). 
	
	CE detects the uncorrelated event $(4;A;started)$, where its TD is checked (cf. Table.~\ref{tbl:TD}). $TD(A)$ is an empty set, which indicates a start of a new case. Hence, case $3$ is added to list of current \emph{Cases}, and event $(4;A;started)$ is now correlated to case $3$. This correlation is trusted 100\% due to the nature of the incoming event as the start of new case. When CE detects the uncorrelated event $(5;A;completed)$, it leads to checking for execution heuristics of the activity $A$. Based on the execution heuristics (cf. Table.~\ref{tbl:heurisitic:data}), an event $(5;A;completed)$ can only be correlated to the started event $(4;A;started)$ in case $3$ with 100\% trust. However, it may differ with other events.
	
	Each event correlation may have different trust percentages; which are assigned based on the expected execution duration of each activity in the process model. The execution duration can be calculated as the difference between current event $e$ and its possible dependencies in the current cases. For example, event $(5;B;started)$ has $TD(B)=\{\{A\},\{N\}\}$, and based on the incoming stream in Table.~\ref{tbl:uncorrelated:2}, the only available dependency is $\{A\}$. The list of possible allocations per current Cases $\{1,2,3\}$ are $\{(1;A;completed),(2;A;completed),(5;A;completed)\}$. The execution durations per each one is $\{5,4,1\}$ respectively. 
	
	An execution duration is calculated as ($end_{e"}$ - $start_e$), where $end_{e"}$ is the timestamp for the incoming event with activity name $e$ (e.g. event $(6;B;completed)$ has ts=6) and $start_e$ represents the completion of one of the correlated dependency event instances listed in the possible allocation list (e.g. $(2;A;completed)$ in case lifecycle attribute has only one value \{completed\} (cf. Table.~\ref{tbl:uncorrelated:1}), or $(5;B;started)$ in case the lifecycle attribute has two values \{started,completed\} (cf. Table.~\ref{tbl:uncorrelated:2})). If TD(event) has a set of concurrent events (e.g. event ($16;L;completed$)), the $start_e$ is replaced with the maximum timestamp of all concurrent events per case. Each duration is checked against the given execution heuristics of an activity, cf. Definition~\ref{def:heuristic:function}, cf. Table.~\ref{tbl:heurisitic:data}. For simplicity, any further reference to events will be shortened to ($ts$;$e$).	
	
	\begin{definition}[Execution Heuristics] Let $A$ be the set of all activities within a process model. Execution Heuristics of an activity \textbf{a} is
		\begin{center}
			\centering	$H = \{(Min_a,Max_a)~|~ a \in A  ~\wedge~ Min_a , Max_a \in \mathbb{R} \}$ 
		\end{center}
		
		where $(Min_a,Max_a)$ is the minimum and maximum execution times for an activity \textbf{a} respectively. Also, $H(a)$ is used to refer to $(Min_a,Max_a)$ pair for activity \textbf{a}. Following a normal distribution of execution, the average execution is calculated as $Avg_a=\lceil\frac{Max_a+ Min_a}{2}\rceil$. Other possible execution times for activity \textbf{a} are represented as $Range_a$, where $Range_a = [Min_a,Max_a]\setminus \{Avg_a\}$.
		\label{def:heuristic:function}		
	\end{definition}
	
	In our example, the list of possible allocations for event $(6;B)$ is filtered based on its execution heuristics $H(B)$. We can find that $H(B)$=(1,4), cf. Table.~\ref{tbl:heurisitic:data}, while the execution durations calculated are $\{5,4,1\}$ for $\{$(1;A), (2;A), (5;A)$\}$ respectively. Hence, event $(1;A)$ is excluded as it is out of heuristics specified. The final list of possible allocations is updated to $\{(2;A),(5;A)\}$, and the list of Cases are $\{2,3\}$ respectively. However, there exists an \emph{uncertainty} about the likelihood of correlating each incoming \emph{event} to a specific case, we employ probabilities to assign event instances probability. 
	
	As each event may be assigned to a case or more with different trust percentages. The trust value is calculated based on \emph{probability} of possible correlation of an event instance $e_i$ within a case $c$, cf. \eqref{eq:event:prob}. We define $S^a_{avg}$ as the list of correlated event instances $e_i$ for an event ($e.activity\_name$=$a$) that have execution duration equivalent to $Avg_a$, either on the same case or over different cases.  This set is updated based on the final list of possible allocations filtered with $TD(a)$ and $H(a)$. A similar list $S^a_{range}$ is specified for the list of correlated event instances that are correlated w.r.t heuristic range $Range_a$. These sets can be based on other probability distributions on the future.
	
	\begin{equation}\label{eq:event:prob}
	p(e_{i}|c)= 
	\begin{cases}
	1 & m =1\\
	\frac {m+1}{m^2} & m >  1 \wedge \forall e_{i}\in S^a_{avg} \\
	\frac{m- \frac{|Avg_a|}{|Range_a|}}{m^2} & m > 1 \wedge \forall e_{i}\in 
	S^a_{range} 
	\end{cases}	
	\end{equation}
	
	where an event $e$ may be cloned into $m$ possible event instances $e_i$ over all cases, $i\in 2, \dots, m$ and $c \in Cases$. Each correlated event instance $e_i$ is either correlated to a case $c$ within average execution time ($Avg_a$), or correlated to a case $c$ within rest of heuristic range ($Range_a$). If an event is only correlated to a single case, then its probability will be 1.
	
	In Table.~\ref{tbl:uncorrelated:1}, event $(13;E)$ has a final list of possible allocations as $\{(7;D),(10;D)\}$ with execution durations $\{6,3\}$ respectively. Both durations are within $H(E)$=(1,7). However event instance $(7;D)$ has $TD(D)$=$\{\{B\}\}$ and $H(D)$=(1,1), and it was correlated after event $(6;B)$ in Cases $\{2,3\}$. While $(10;D)$ was correlated after event $(9;B)$ in case~3. Hence, case 2 has one possible allocation for event ($7;D$), while case 3 has two possible allocations for the same event. This will need the help of probabilities in calculating the trust percentage for each event correlation per a case, cf. Definition~\ref{def:event:trust}.

	%%%%%%%%%%%%%%%%%%%%%%%%%%%%%%%%%%%%%%%%%%%%%%%%%%%%%%%%%%%%%%%%%%%%%%%%%%%%%%%%%%%%%%%

	\begin{definition} [Event Correlation Trust]
		\label{def:event:trust}	
		Let $p(e_{i}|c)$ represents correlation probability to case $c$ for an event $e$ (cf. \eqref{eq:event:prob}). The correlation of an event $e$ to its respective case $c$ has (\textbf{trust}) value calculated as  
		\begin{center}
			\centering $trust(e_{i},c) = 100 \times (\sum^m_{i=1} p(e_{i}|c)) $ \%
		\end{center}
		
		where $e_{i}.activity\_name \in$ Activities, $m$ is the number of event instances $e_i$ that are correlated per case $c \in Cases$, and $c = \{x | x\in CorrelatedEvents\}$. \textit{Activities} is the set of $activity\_names$ from the process model, and \textit{Cases} list is initiated and updated with each incoming event $e$, where TD(e)=\{\}. 
		
	\end{definition}
	
	An event may be correlated to different cases, each one of them has its event probability w.r.t its case. However, if the trust percentage of an event $<100\%$ or occurs in a cyclic trace of the model, then multiple occurrences can take place in the same case with different trust percentages. For example, an event $E$ can occur in cyclic route after either dependencies $\{D\}$ or $\{H\}$. In some scenarios, an event is correlated to only one case $c$, with correlation trust equals 100\% as with event $(5;A)$ in case 3. However, in other scenarios, an event $e$ may fail to correlate with any case $c \in Cases$, due to either inaccurate execution heuristics, or noisy events~\cite{VanderAalst2004}. Hence, its trust value equals zero. Any uncorrelated events can be the main reason for the deviations in the process model and thus non-conforming model. Also, it can be the main source of violations in compliance monitoring.

	%%%%%%%%%%%%%%%%%%%%%%%%%%%%%%%%%%%%%%%%%%%%%%%%%%%%%%%%%%%%%%%%%%%%%%%%%%%%%%%%%%%%%%%%
	
	Algorithm~\ref{algo:CE} is triggered by each incoming event $e$ in the stream of uncorrelated events. It uses the generated task dependencies $TD$ from Algorithm~\ref{algo:build:TD}, and the execution heuristics $H$ (cf. Definition~\ref{def:heuristic:function}) as input parameters. Line~\ref{line:CE:new:case} is a special scenario where an event $e$ represents a start of new case $caseID$ with 100\% trust. In a more general setting, multiple checks are required to correlate events to their respective cases. First, it searches for any possible cases satisfying $TD$ of the incoming event $e$, see line~\ref{line:CE:case:TD}. Then, the set of possible cases are filtered using execution heuristics (cf. \eqref{eq:event:prob}), see line~\ref{line:CE:case:H}. 
	
	To finalize the set of possible allocations for an event $e$, one more check is performed. Line~\ref{line:CE:no:loop} checks if $isLoopEntry$ is false, then exclude all cases containing an occurrence of the same event with trust 100\%. Otherwise, this event can occur multiple times in the same case. If there are no possible allocations, the event $e$ fails to correlate to any case and has 0\% trust, which indicates a deviation from the original model, see line~\ref{line:CE:no:correlation}. Otherwise, for each candidate D, a new instance of the correlated event $e$ is added with trust\% (cf. Definition~\ref{def:event:trust}), see line~\ref{line:CE:yes:correlation}. Finally, the correlated event instances $e_i$ are inserted into \emph{CorrelatedEvents} set (cf. Definition~\ref{def:CorrelatedEvents}), see line~\ref{line:CE:insert}.

	%%%%%%%%%%%%%%%%%%%%%%%%%%%%%%%%%%%%%%%%%%%%%%%%%%%%%%%%%%%%%%%%%%%%%%%%%
	\begin{algorithm}[bt]\caption{Correlate Events}\label{algo:CE}
		\scriptsize
		\begin{algorithmic}[1]
			\Require $e(ts,name)$ ~~//uncorrelated event (Table.\ref{tbl:uncorrelated:1}),$ts$: timestamp,$name$: event name
			\Require $TD$  ~~~~~~~~~~~~~//the task dependencies (Table.\ref{tbl:TD})
			\Require $H$ ~~~~~~~~~~~~~~~~//the heuristics about activity executions  (Table.\ref{tbl:heurisitic:data})
			\Ensure \emph{CorrelatedEvents} ~~//updated with correlated event instances (Definition\ref{def:CorrelatedEvents})
			%		\State $Activities$ = all activities specified in Process Model.
			\State $Cases$ = all cases generated from CE ~~~//updated with each detected new case.
			\State $PAlloc$ = $\{\}$~~~~ // List of possible allocations
			\State $PCase$ = $\{\}$~~~~ // List of possible cases
			\State $Dependents$ = $TD(e.name)$~~//Task Dependency of event $e$ (cf.  Definition\ref{def:TaskDependencies})
			\If{($Dependents=\{\}$)}~~~~~~~~~~~~~~~~~~~~~~ //add new case $c$ \label{line:CE:new:case}
			\State $caseID=|Cases|+1$
			\State $CorrelatedEvents.insertInstance(ts,name,caseID,100\%)$
			\Else
			\ForAll {($D\in Dependents$)}
			\If{($|D|==1 $)} 
			\State	$PCase = Cases.contains(D)$\label{line:CE:case:TD}
			\State 
			$PAlloc = PAlloc \bigcup getInstances(PCase , H(e))$ \label{line:CE:case:H} %(cf. Fig.\ref{fig:AND:query})
			\ElsIf {
				($|D|>1$)} //searches CorrelatedEvents for possible cases 
			\ForAll{($x\in D$)} %~~~//for concurrent activities $x$ in $D$
			\State $PCase = PCase\bigcup Cases.contains(x.activity\_name)$
			\If{( $\neg(isLoopEntry(x))$)} \label{line:CE:no:loop}
			\State \begin{varwidth}[t]{\linewidth}$excludeCases=PCase.contains(x.activity\_name,100\%)$~~\\//exclude cases having \emph{x} with trust 100\% \end{varwidth}\vspace{1mm}
			\State $PCase = PCase~\setminus ~excludeCases$
			\EndIf
			\State 
			$PAlloc=PAlloc \bigcup  getInstances(PCase,H(e))$
			\EndFor
			\EndIf
			\EndFor
			\If {($|PAlloc|==0$)} \label{line:CE:no:correlation}  //cannot be assigned to any existing case
			\State $caseID = \bot$;  
			\State $trust = \bot$
			\State $CorrelatedEvents.insertInstance(ts,name,caseID,trust)$
			\Else
			\ForAll{($e_i \in PAlloc$)} \label{line:CE:yes:correlation}
			\State $caseID = e_i.caseID$
			\State $trust = trust(e_i,~caseID) $ ~~//using Definition\ref{def:event:trust} \label{line:CE:insert} %eq:\ref{eq:event:prob} in 
			\State $CorrelatedEvents.insertInstance(ts,name,caseID,trust)$
			\EndFor
			\EndIf		
			\EndIf
		\end{algorithmic}
	\end{algorithm}
	%%%%%%%%%%%%%%%%%%%%%%%%%%%%%%%%%%%%%%%%%%%%%%%%%%%%%%%%%%%%%%%%%%%%%%%%%

	Table.~\ref{tbl:dashboard-snapshot} represents a snapshot of the output of our approach. It displays events and their correlated \emph{case ID} with different trust percentage (cf. Definition~\ref{def:event:trust}). For example, event $(16;L)$ has $TD(L)=\{\{G\},\{\{I, J\}\}$, where either both activities $I$ and $J$ must occur before the event $L$, or activity $G$ happen before the event $L$ in a case, given that they satisfy the heuristics condition of $(16;L)$. This is only applicable for cases $1$ and $2$. 
	
	\begin{table}[bt]
		\caption{Snapshot of Correlated Events Stream}
		\label{tbl:dashboard-snapshot}
		\def\arraystretch{1}
		\ignorespaces 
		\scriptsize
		\begin{center}
			\begin{tabular}{cccc}
				\hline
				\textbf{Case ID} & \textbf{Timestamp} & \textbf{Activity} & \textbf{Trust \%} \\ 
				\hline
				\dots & \dots &	\dots & \dots \\
				3 & 2019-6-16 11:55:05 & A & \textbf{100} \\
				2 & 2019-6-16 11:55:06 & B & 50 \\
				3 & 2019-6-16 11:55:06 & B & 50 \\
				2 & 2019-6-16 11:55:07 & D & 50 \\
				3 & 2019-6-16 11:55:07 & D & 50 \\
				1 & 2019-6-16 11:55:08 & J & 50 \\
				2 & 2019-6-16 11:55:08 & J & 50 \\
				3 & 2019-6-16 11:55:09 & B & \textbf{100} \\
				3 & 2019-6-16 11:55:10 & D & \textbf{100} \\
				1 & 2019-6-16 11:55:11 & I & 50 \\
				2 & 2019-6-16 11:55:11 & I & 50 \\
				2 & 2019-6-16 11:55:13 & E & 33.34 \\
				3 & 2019-6-16 11:55:13 & E & 66.67 \\
				\dots & \dots &	\dots & \dots \\
				1 & 2019-6-16 11:55:16 & L & 50 \\
				2 & 2019-6-16 11:55:16 & L & 50 \\
				\dots & \dots &	\dots & \dots \\
				\hline
			\end{tabular}%
		\end{center}
	\end{table}
	
	\begin{table}[bt]
		\caption{Correlated Events Stream for Case 2}
		\label{tbl:dashboard-case2}
		\def\arraystretch{1}
		\ignorespaces 
		\scriptsize
		\begin{center}
			\begin{tabular}{cccc}
				\hline
				\textbf{Case ID} & \textbf{Timestamp} & \textbf{Activity} & \textbf{Trust \%} \\ 
				\hline
				2 & 2019-6-16 11:55:02 & A & 100.00 \\
				2 & 2019-6-16 11:55:03 & B & 50.00 \\
				2 & 2019-6-16 11:55:04 & C & 50.00 \\
				2 & 2019-6-16 11:55:06 & B & 50.00 \\
				2 & 2019-6-16 11:55:07 & D & 50.00 \\
				2 & 2019-6-16 11:55:08 & J & 50.00 \\
				2 & 2019-6-16 11:55:11 & I & 50.00 \\
				2 & 2019-6-16 11:55:13 & E & 33.33 \\
				2 & 2019-6-16 11:55:14 & E & 27.78 \\
				2 & 2019-6-16 11:55:15 & F & 50.00 \\
				2 & 2019-6-16 11:55:16 & L & 50.00 \\
				2 & 2019-6-16 11:55:17 & G & 50.00 \\
				2 & 2019-6-16 11:55:18 & H & 50.00 \\
				2 & 2019-6-16 11:55:19 & E & 50.00 \\
				2 & 2019-6-16 11:55:20 & G & 50.00 \\
				2 & 2019-6-16 11:55:21 & L & 50.00 \\
				2 & 2019-6-16 11:55:22 & L & 50.00 \\
				2 & 2019-6-16 11:55:23 & N & 33.33 \\
				2 & 2019-6-16 11:55:24 & B & 50.00 \\
				2 & 2019-6-16 11:55:25 & M & 37.50 \\
				2 & 2019-6-16 11:55:26 & C & 50.00 \\
				2 & 2019-6-16 11:55:27 & I & 50.00 \\
				2 & 2019-6-16 11:55:28 & M & 33.33 \\
				2 & 2019-6-16 11:55:29 & J & 50.00 \\
				2 & 2019-6-16 11:55:31 & L & 50.00 \\
				2 & 2019-6-16 11:55:32 & M & 33.33 \\
				\hline
			\end{tabular}%
		\end{center}
	\end{table}

	The \textit{CorrelatedEvents} are also stored on an offline log as a snapshot of the full execution of the REC module. This log can be further analyzed by different applications, such as \emph{predictive} analysis for compliance management~\cite{Mulo2013,Barnawi2016}, or \emph{automating} decision making~\cite{Kemsley2015,Doganata2012}. Table.~\ref{tbl:dashboard-case2} presents the whole set of correlated events for Case 2, it illustrates the cyclic behavior of the model \{$\{E\},\{F\},\{H\},\{E\}$\} as well as parallel execution of events \{\{$I$, $J$\}\}. These representations can be useful for an online conformance checking mechanism.

	\section{Results and Discussion} \label{sec:results:discussion}
	
	In this section, we discuss the implementation of our approach and the REC algorithms in a real life setting based on real logs from BPI challenges~\cite{Buijs2014,Steeman2013} to assess the applicability of our approach. 
	
	\subsection{REC implementation}\label{sub:sec:evaluation:implementation}
	
	\begin{lstlisting}[
	language=SQL,
	deletekeywords={IDENTITY},
	deletekeywords={[2]INT},
	morekeywords={clustered},
	framesep=1pt,
	xleftmargin=30pt,
	framexleftmargin=30pt,
	frame=tb,
	framerule=0pt,
	basicstyle=\scriptsize ]
	SELECT 'timestamp', activity_name, caseID, trust, diffTS  
	FROM ( SELECT 'timestamp', activity_name, caseID, trust, (STRFTIME('%s','2019-06-16 11:55:09')-STRFTIME('%s','timestamp')) AS diffTS  
	FROM CorrelatedEvents 
	WHERE caseID !=NULL AND ( activity_name='A' OR ( activity_name='N'  AND caseID IN (  
	SELECT caseID  FROM CorrelatedEvents 
	WHERE activity_name='A' AND caseID NOT IN ( 
	SELECT caseID  FROM CorrelatedEvents 
	WHERE activity_name='B' AND trust= 100.0 AND 'timestamp' > (SELECT MAX('timestamp') FROM CorrelatedEvents WHERE activity_name='A' GROUP BY caseID)  AND 'timestamp' < (SELECT MIN('timestamp') FROM CorrelatedEvents WHERE activity_name='N' GROUP BY caseID)  
	) ) ) ) )AS x 
	WHERE diffTS >= 1.0 AND diffTS <= 4.0  ORDER BY caseID ;
	\end{lstlisting}

	The previous script illustrates a sample query of how event correlation filtration is done. The query checks for possible cases of event B, its timestamp `2019-06-16 11:55:09', that happens within its heuristics, cf. Tables.~\ref{tbl:heurisitic:data},\ref{tbl:uncorrelated:1}. First, it checks the task dependency of event B; TD(B) has dependency with exclusive events A and N, where isLoopEntry(A) = False and isLoopEntry(N) = True. Then, it checks its execution heuristics; it ranges from 1 to 4 sec. Finally, it checks for previous occurrence of B in these cases after A with trust 100\%. This query gets all possible allocations of event instances satisfying dependency of event (9,B) (i.e. either events A or N), within the specified execution heuristics.

	Complexity of building Task Dependencies, cf. Algorithm~\ref{algo:build:TD}, is O(km), where m is the number of activities in the process model, and k is the number of activities with exclusive dependencies; where k \ensuremath{\in } [1,m]. A further checking is performed to detect possible loop entries. While, the complexity of building query time in Correlate Events, cf. Algorithm~\ref{algo:CE} is O(kn), where n is the count of possible instances in set of Activities that an event e can correlate to their possible Cases, and k is the number of dependency activities for that event based on TD, cf. Definition~\ref{def:TaskDependencies}. 
	
	The performance of REC can only be affected by the speed of incoming event in the uncorrelated stream of events S. The Correlate Events component takes up to $\approx$20 milliseconds as a processing time for each incoming event. Moreover, the accuracy of correlation is highly dependent on the correctness of execution heuristics specified. Missing or incorrect results may happen based on how narrow or broad the execution duration of an event respectively.   
	
	We evaluate REC against Runtime Deduction of Case ID (RDCI), and Deducing of Case Identifier cyclic (DCIc) approaches \cite{Helal2015,BayomieAE16}. Both RDCI and DCIc are based on case decision trees which grow exponentially with the number of incoming events. REC produces correlated events stream at almost the same time of their occurrence using database indexing. RDCI is sensitive to shallow trees with multiple cases, which is irrelevant for REC. Also, it does not support cyclic models. While both REC and DCIc support them. 
	
	Table.~\ref{tbl:evaluation:tech:REC} illustrates the differences between REC and similar approaches. Both REC and DCIc techniques are supporting cyclic models, while RDCI, Expectactation-Maximization (E-Max)~\cite{Ferreira2009}, and Correlation Miner~\cite{Pourmirza2015} are only applied on acyclic models. REC and RDCI are correlating at runtime, while DCIc, E-Max, Correlation Miner correlate events offline. Moreover, REC, RDCI, and DCIc techniques share same input requirements as well as their sensitivity to the accuracy of execution heuristics. E-Max only needs the process model for correlation, while Correlation Miner takes the process model along with mapping constraints to generate a correlated log to help in producing a mined orchestration model.

	\begin{table}[bt]
		\caption{Comparing different techniques for correlating events}
		\label{tbl:evaluation:tech:REC}
		\def\arraystretch{1}
		\ignorespaces 
		\scriptsize
		\begin{center}
			\setlength\tabcolsep{2pt} %Default is 6pt for horizontal spacing
			\begin{tabular}{|c|c|c|c|c|}
				\hline
				\textbf{Technique}
				& \textbf{Input}
				& \textbf{Output}
				& \textbf{Acyclic}
				& \textbf{Cyclic} \\ 
				\hline
				
				REC
				& \multirow{3}{*}{\begin{tabular}{l}Process Model+\\Uncorrelated Log+ \\Heuristics (sensitive)\end{tabular}}
				& \multirow{2}{*}{\begin{tabular}{l}Online Correlated \\Events Stream\end{tabular}} 
				& +
				& + \\ 
				\cline{1-1}
				\cline{4-5}
				%		 \hline
				RDCI  \cite{Helal2015}
				& 
				& 
				& +
				& - \\
				\cline{1-1}
				\cline{3-5} 
				%		 \hline
				DCIc  \cite{BayomieAE16}
				& 
				& Offline Correlated Event Logs
				& +
				& + \\ 
				\hline
				
				E-Max \cite{Ferreira2009}
				& Uncorrelated Log
				& Offline Correlated Event Log
				& +
				& - \\
				\hline
				
				\begin{tabular}{c}Correlation\\Miner \cite{Pourmirza2015}\end{tabular}
				& \begin{tabular}{l}Uncorrelated Log+\\Mapping Constraints\end{tabular}
				& Mined Orchestration Model
				& +
				& - \\
				\hline
			\end{tabular}
			
		\end{center} 
	\end{table}

	%%%%%%%%%%%%%%%%%%%%%%%%%%%%%%%%%%%%%%%%%%%%%%%%%%%%%%%%%%%%%%%
	\subsection{Evaluation Procedure}
	\label{sub:sec:evaluation:procedure}

	\begin{figure}[bt]
		\centerline{\includegraphics[width=0.5\textwidth]{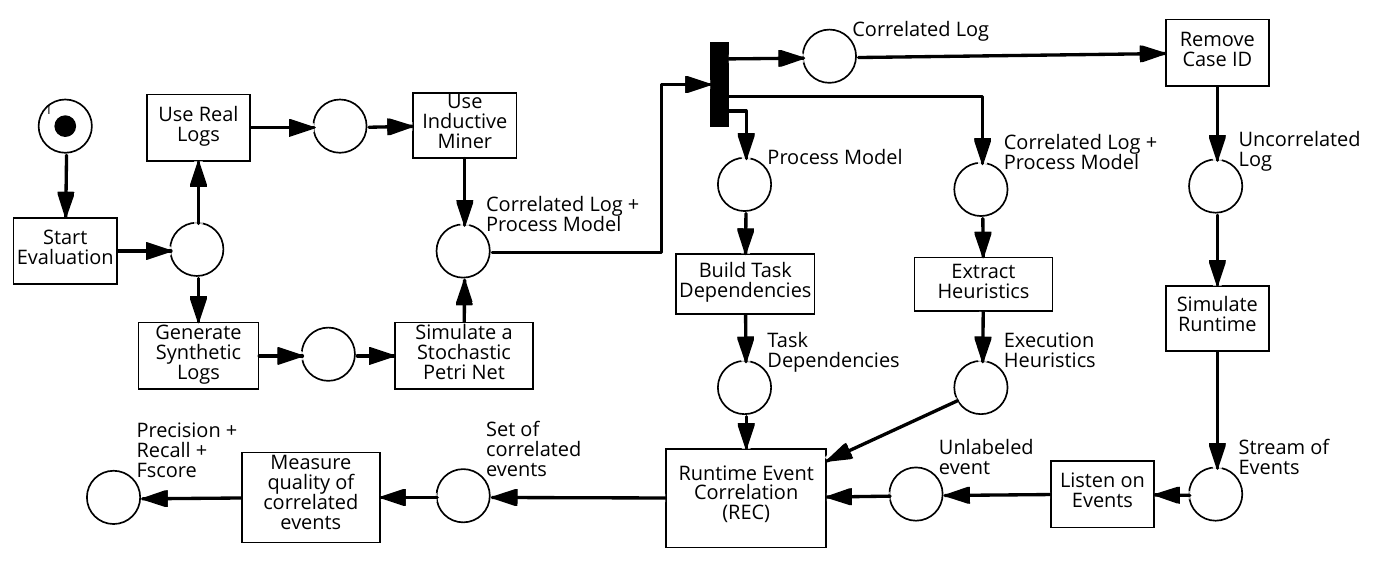}}
		\caption{{REC Evaluation steps}}
		\label{fig:evaluationSteps}
	\end{figure}

	Fig.~\ref{fig:evaluationSteps} shows the evaluation steps of REC with both synthetic and real life logs~\cite{Buijs2014,Steeman2013}. There are two possible scenarios while evaluating this approach: 1) Generate synthetic logs, using the ProM plug-in~\cite{VerbeekBDA10}: ``Perform a simple simulation of a (stochastic) Petri net''~\cite{Rogge-Solti2013b}. The simulated log is updated to reflect the heuristic data. 2) Use real life logs, using the ProM plug-in: ``Mine Petri net with Inductive Miner''~\cite{Leemans2014} to obtain the process model. Then, we extract heuristic information from the real-life log using a tool we built considering business logic. In either case, we remove caseID from the correlated log to produce an uncorrelated log. Also, we build the TD for the process model.
	
	Table.~\ref{tbl:evaluation:tech:REC} compares the accuracy of execution of REC, RDCI and DCIc as presented in~\cite{BayomieAE16} on real life event logs~\cite{Buijs2014,Steeman2013} as well as synthetic log. These measures are presented in~\cite{Doganata2012}, where precision (or specificity \ensuremath{\theta }) is calculated as:  
	
	\begin{equation}
	\theta = TP / (TP + FP)
	\label{eq:precision}
	\end{equation}
	
	Recall (or sensitivity $\eta$) is calculated as:
	\begin{equation}
	\eta = TP / (TP + FN)
	\label{eq:recall}
	\end{equation}
	where, TP represents the number of events correctly correlated, FP is the number of events incorrectly correlated, and FN is the number of events that failed to correlate.
	
	Finally, F-score is calculated as 
	\begin{equation}
	F-score = (2 * \theta * \eta) / (\theta + \eta)
	\label{eq:Fscore}
	\end{equation}

	\begin{table*}[bt]
		\caption{Comparing performance of different approaches for different logs \cite{Helal2020}}
		\label{tbl:evaluation:result:REC}
		\def\arraystretch{1}
		\ignorespaces
		\scriptsize 
		\begin{center}
			\begin{tabular}{|c|c|l|l|l|l|l|l|l|}
				\hline 
				Data set  
				& Basics
				& Approach
				& Precision
				& Recall
				& F-score
				& Execution time
				& Simulation time
				& Average execution per event
				\\ 
				\hline
				
				\begin{tabular}{c}CoSeLoG \\ \cite{Buijs2014}\end{tabular}
				&\begin{tabular}{l}1434 cases\\8556 events\end{tabular}
				& \begin{tabular}{l}REC\\RDCI\\DCIc\end{tabular}
				& \begin{tabular}{l}0.8166\\0.0067\\0.82\end{tabular} 
				& \begin{tabular}{l}0.7822\\0.3588\\0.72\end{tabular} 
				& \begin{tabular}{l}0.799\\0.0132\\0.7668 \end{tabular} 
				& \begin{tabular}{l}$\approxeq$1 min\\$\approxeq$1 min\\35 min\end{tabular}  %$\approxeq$ 43200 sec
				& \begin{tabular}{l}$\approxeq$1 min\\$\approxeq$1 min\\--\end{tabular}
				& \begin{tabular}{l}7.0126 ms\\4.0105 ms\\--\end{tabular}
				\\ 
				\hline
				
				\begin{tabular}{c}BPI2013 \\ \cite{Steeman2013}\end{tabular}
				&\begin{tabular}{l}1487 cases\\6660 events\end{tabular}
				& \begin{tabular}{l}REC\\RDCI\\DCIc\end{tabular}
				& \begin{tabular}{l}0.9985\\0.4474\\0.81\end{tabular} 
				& \begin{tabular}{l}0.9984\\0.0016\\0.72\end{tabular} 
				& \begin{tabular}{l}0.9984\\0.0032\\0.7624\end{tabular} 
				& \begin{tabular}{l}$\approxeq$7 min\\$\approxeq$7 min\\40 min\end{tabular}  
				& \begin{tabular}{l}$\approxeq$7 min\\$\approxeq$7 min\\--\end{tabular}     
				& \begin{tabular}{l}63.0631 ms\\63.0631 ms\\--\end{tabular}
				\\ 
				\hline
				
				\begin{tabular}{c}Synthetic \\(cyclic)\end{tabular}
				&\begin{tabular}{l}1000 cases\\8537  events\end{tabular}
				& \begin{tabular}{l}REC\\RDCI\\DCIc\end{tabular} 
				& \begin{tabular}{l}0.8287\\0.0369\\0.88\end{tabular} 
				& \begin{tabular}{l}0.9902\\0.0369\\0.79\end{tabular} 
				& \begin{tabular}{l}0.9023\\0.0342\\0.8326\end{tabular} 
				& \begin{tabular}{l}$\approxeq$10 min\\$\approxeq$10 min\\$\approxeq$3 hr\end{tabular}  
				& \begin{tabular}{l}$\approxeq$10 min\\$\approxeq$10 min\\--\end{tabular}     
				& \begin{tabular}{l}7.4559 ms\\3.0508 ms\\--\end{tabular}
				\\ 	\hline
				
			\end{tabular}
		\end{center}
		
	\end{table*}

	The CoSeLoG data set~\cite{Buijs2014} has relatively less f-score ($\approx$0.77) than the other data sets for both REC and DCIc techniques. These results are based on the complexity of the original process model. CoSeLoG has about six activities with self-loops, each one of them has a high frequency of occurrence in the original log, while both BPI2013~\cite{Steeman2013} and Synthetic logs have two activities only with self-loops. In the BPI2013 data set, the activities with self-loops have relatively higher occurrence frequency than the Synthetic log. This highly affects the f-score for DCIc (ranges from 0.76 to 0.83), while REC is not as affected with how frequent a self-loop takes place (f-score {\textgreater} 0.9). 
	
	The other factor to consider while comparing the REC, the RDCI, and the DCIc techniques is the time. The total execution time expresses the time needed to correlate the events to their case identifier as well as the simulation time (if available). The simulation time is calculated for the techniques applied in an online setting, i.e. RDCI and REC. The DCIc technique is applied in an offline setting.

	The simulation time represents the amount of time to rerun the system in a faster speed than the original models; i.e. to replay the system generation of events. For example, the CoSeLoG data set spans over $\approx$ a year and three months, while its simulation time takes $\approx$1 minute. Also, the BPI2013 data set spans over $\approx$2 years, while its simulation time takes $\approx$7 minutes. The Synthetic log execution span is $\approx$ a day, and its simulation time takes $\approx$10 minutes. Moreover, each data set used in the evaluation has a different nature of time frequency, e.g. days, minutes, seconds. For example, CoSeLoG has time frequency of minutes, while BPI2013 has time frequency of days. Finally, the Synthetic log has time frequency of seconds.

	The last column in the comparison expresses the average execution time for processing each incoming event. This number had two factors affecting it. The first is the total time of execution, and the second is how frequent an event is detected. In BPI2013, some events wait up to 7 months to occur. Hence, the average execution time is affected by the waiting time in the simulation as well as the processing time. Both the CoSeLoG data set and Synthetic log have an average execution time $\approx$7 milliseconds. Moreover, the REC correlation accuracy is sensitive to the correctness of the execution heuristics. The narrower the heuristics ranges, the more the approach fails to correlate events correctly. While the wider are the heuristics ranges, the more incorrect are the correlated events.

	On the other hand, other real-life logs, BPI2017~\cite{VanDongen2017} and Sepsis cases\unskip~\cite{Sepsis2016}, have generated inaccurate results, F-score=0.56578 and 0.8695 respectively. The main factor affecting the accuracy of our model in both logs is the number of loops in the original process models. The increase in the number of loops worsens the accuracy dramatically as in BPI2017. The original log for BPI2017 has 1,202,267 events for only 31509 cases. However, a sample of the log was tested, 1836 events for 100 cases. BPI2017 had many cyclic behaviors in the original traces, which affected the trust percentage of each correlated event. Moreover, the original process model for Sepsis cases had a cyclic behavior for the set of start events, which is not supported in our approach. This has affected the resulted number of cases drastically (10168 cases instead of 1050 cases). REC assumes that there is no loop behavior for the set of starting events.

	\section{Conclusion and Future work}\label{sec:conclusion:future}
	
	In this paper, we introduced a Runtime Event Correlation (REC) approach for unmanaged events. It correlates a stream of uncorrelated events to their respective cases at runtime, using a database storage, SQLite. We use some additional inputs in a process-aware model to label the events into a stream of correlated events with different trust percentages. We take as inputs: a Petri net process model, the execution heuristics about each activity in the process model, in addition to the stream of uncorrelated events.
	
	REC detects and observes an uncorrelated event in near real-time and provide an immediate response to the set of correlated event instances with different trust percentages. If the observed event was considered noisy as defined in~\cite{ProcessMiningBook2016}, it is directly mentioned on the stream of correlated events. Also, REC can address the incompleteness of the event log~\cite{ProcessMiningBook2016}, i.e. a snapshot from a process execution, which violates the process model or deviation of the original process model, since each new detected uncorrelated event is considered part of an incomplete event log.
	
	One of the main advantages of our approach is supporting both acyclic and cyclic models, either structured or arbitrary loops. Also, the execution performance of correlation process is almost same as the originally simulated real-life logs. However, it is affected by the speed of incoming events, as each event needs up to $\approx$20 milliseconds of processing which is a challenge in real-time streaming of events. Also, our accuracy of event correlation is affected by the accuracy of execution heuristics. If the heuristics are incorrectly specified, an erroneous correlated event is highly expected. 
	
	As a future work, our approach can be migrated and tested for larger systems and more critical ones using big data and the cloud for storage and processing. Considering the usage of any in-memory database engines, applying near real-time monitoring of real systems can also be very interesting. Moreover, expanding the usage of correlated event logs to other applications, e.g. discovery and enhancements can be quite challenging.
	
	Other extensions can be added such as 1) repair events at runtime to complete other missing information; such as: timestamp, resources, secondary identifiers, etc., 2) correlate events at runtime for middle quality event streams, i.e. consider the full business process life cycle, the performing resources and roles, etc., 3) specify a criteria for finding a correct execution heuristics. Finally, our approach can be migrated with compliance monitoring frameworks and conformance checking approaches to overcome the low-quality of the logs and get better results.

%	\section*{References}

%	\bibliographystyle{IEEEtran}
%	\bibliography{IEEEabrv,references3}
		
\end{document}